\documentclass[a4paper,12pt]{article}

\usepackage{latexsym}

\newcommand{\ud}{\mathrm{d}}

\begin{document}

\title{Metric gravity theories and cosmology.\\
II. Stability of a ground state in $f(R)$ theories}

\author{Leszek M. SOKO\L{}OWSKI \\
Astronomical Observatory and Centre for Astrophysics,\\
Jagellonian University,\\ 
Orla 171,  Krak\'ow 30-244, Poland \\uflsokol@th.if.uj.edu.pl} 

\date{}
\maketitle

\centerline{Short title: Nonlinear gravity in cosmology}

\begin{abstract}
In the second part of investigation of metric nonlinear gravity theories we 
study a fundamental criterion of viability of any gravity theory: 
the existence of a stable ground state solution being either Minkowski, de 
Sitter or anti--de Sitter space. Stability of the ground state is independent 
of which frame is physical. In general a given theory has multiple ground 
states and splits into independent physical sectors. The fact that all 
$L=f(g_{\alpha\beta}, R_{\mu\nu})$ gravity 
theories (except some singular cases) are dynamically equivalent to Einstein 
gravity plus a massive spin--two and a massive scalar field allows to 
investigate the stability problem using methods developed in general 
relativity. These 
methods can be directly applied to $L=f(R)$ theories wherein the spin--two 
field is absent. Furthermore for these theories which have anti--de Sitter 
space as the ground state we prove a positive--energy theorem allowing to 
define the notion of conserved total gravitational energy in Jordan frame 
(i.e., for the fourth--order equations of motion). As is shown in thirteen  
examples of specific Lagrangians the stability criterion works effectively 
without long computations whenever the curvature of the ground state is 
determined. An infinite number of gravity theories have a stable ground 
state and further viability criteria are necessary. 
\end{abstract}


\section{Introduction} 
In our previous work \cite{Sok} (hereafter cited as Paper I) we have argued 
that the cosmological observations based on the spatially flat 
Robertson--Walker spacetime (usually fitted by the $\Lambda$CDM model) are 
unsuitable and insufficient to reconstruct the Lagrangian of the true 
gravity theory which correctly accounts for the present epoch of cosmic 
acceleration. The very fact that the observational data may be fitted by 
a huge collection of diverse Lagrangians clearly indicates that the idea of 
reconstructing the correct theory from cosmology is implausible and a 
deeper investigation confirms that the task is impossible. If one believes 
that general relativity needs some modifications, these should not be 
directly induced from the approximate data. Rather, as any other new 
physical theory, a new gravity theory should be based on new concepts and 
ideas. In the present case the modifications consist in replacing the 
Einstein Lagrangian $L=R$ by a nonlinear function of Riemann tensor and it 
is difficult to give a physical idea which would uniquely choose the 
correct function. One should rather apply various criteria taken from 
classical field theory to maximally narrow down the class of viable 
theories. In our opinion the most fundamental condition is the existence 
of a maximally symmetric stable ground state solution to the equations of 
motion in pure gravity. Ordinary matter cannot destroy stability of a  
ground state if it is stable against purely gravitational perturbations. 
Stability of the ground state is independent of which frame (the set of 
dynamical field variables) is regarded as physical. Hence the stability 
can be investigated in Einstein frame where the most reliable methods of 
checking it have been developed. \\
In the case of $L=f(R)$  the stability is verified in this frame in a 
very effective and 
quick way. The main step is to solve an algebraic equation to determine 
the curvature $R$ of a ground state. If the equation is solvable for a 
given Lagrangian one easily establishes whether the state is stable. It 
is worth noting that our method based on the dominant energy condition 
for the scalar component of gravity, applies to generic (inhomogeneous) 
perturbations of a ground state, while most authors assume homogeneous 
or at most spherically symmetric perturbations.\\

We emphasize that stability of the ground state and other possible 
viability conditions concern physical viability of a gravity theory 
from the viewpoint of field theory, i.e.~they concern its internal 
structure and its relationships to other physical theories. There are 
many viable gravity theories. At the present level of knowledge there 
is no system of selection rules (i.e.~viability conditions) allowing one 
to uniquely determine the correct theory. Hence physical viability has 
a restricted meaning and does not mean that a viable theory 
necessarily fits in a 
satisfactory way some empirical data. Actually most of viable gravity 
theories is in conflict with observations. In particular the physical 
viability (which in the present work coincides with the stability of 
the ground state) is independent of a cosmological viability 
introduced by Amendola et al. \cite{AGPT}. According to these authors 
the cosmological viability means a satisfactory evolution of the 
universe in the flat ($k=0$) Robertson--Walker spacetime: long matter 
era with the cosmic scale factor $a\propto t^{2/3}$ prior to a 
late-time acceleration epoch.\\

The present work deals with various aspects of the stability problem: 
determination of ground state solutions and their multiplicity, notion 
of stability, total energy and its relationship to stability of a 
ground state, reliability of various methods of checking the stability. 
Finally we formulate a stability condition in terms of a potential in 
Einstein frame for the scalar component of a $L=f(R)$ gravity. We apply 
the condition to thirteen specific Lagrangians, mainly taken from the 
literature and show\footnote{The problem of the cosmic acceleration 
applying $f(R)$ gravity was also investigated within the Kaluza--Klein 
framework \cite{BRu} and some results achieved there are akin to ours.} 
that some theories which are cosmologically viable are physically 
untenable. 

\section{Stability of a maximally symmetric ground state}
A minimal requirement that may be imposed on a gravity theory to be 
viable  is 
that it has a classically stable maximally symmetric ground state solution. 
In some classical field theories, e.g.~in Liouville field theory 
\cite{dHJ} a ground state may not exist, but in gravitational physics the 
existence of a ground state hardly needs justification. In a metric gravity 
theory gravitational interactions are manifested by the dynamical 
curvature of the 
spacetime, hence in the absence of these interactions the spacetime 
should be either flat or maximally symmetric with the nongeometric 
components of the gravitational multiplet equal to zero or covariantly 
constant. Therefore the spacetime of the ground state for any NLG theory 
may be Minkowski, de Sitter (dS) or anti--de Sitter\footnote{By anti--de 
Sitter space we always mean the covering anti--de Sitter space without 
closed timelike curves.} (AdS) space. For simplicity we assume spacetime 
dimensionality 
$d=4$ although our arguments (with slight modifications) will also hold 
in $d>4$. Classical stability means that the ground state solution is 
stable against small excitations of the (multicomponent) gravitational 
field and small excitations of a given kind of matter sources, i.e.~there 
are no growing in time perturbation modes. In principle a viable classical 
field theory may admit a semiclassical instability: the ground state is 
separated by a finite barrier from a more stable (in the sense of lower 
energy) state and can decay into it by a semiclassical barrier penetration 
\cite{Wi}. We shall not consider this possibility and focus our attention 
on classical stability, hereafter named stability. \\

A question that may arise at the very beginning of investigation of the 
problem is whether a metric NLG theory, being a higher derivative one, can 
at all be stable \cite{Woo}. In point particle mechanics one may invoke to 
this end the old famous Ostrogradski theorem to the effect that if a 
mechanical Lagrangian depends on second and higher time derivatives of 
the particle positions (which cannot be eliminated by partial integration) 
the corresponding Hamiltonian is linear in at least one canonical momentum 
and thus is unbounded from below. As a consequence there are both positive 
and negative energy states and if the particles are interacting the theory 
is unstable since any solution decays explosively due to self--excitation: 
unlimited amount of energy is transferred from negative energy particles 
to positive energy ones. By analogy, the same (or rather more drastic and 
violent) instability is expected to occur in classical (and quantum) field 
theory with higher time derivatives. Thus a generic NLG theory should be 
inherently unstable and hence unphysical. We admit that the problem is 
important and deserves a detailed investigation. Here we wish only to make 
short comment on how it is possible to avoid this conclusion. \\
We stress that the Ostrogradski theorem is a rigorous "no--go theorem" in 
classical and quantum point particle mechanics \cite{Woo} while in metric 
NLG theories it may only be conjectured by analogy. In fact, a mechanical 
Hamiltonian determines energy and if it is indefinite (and unbounded from 
below) it signals that self--excitation processes are likely to occur.  
Recall that a metric theory of gravity is based on the equivalence principle 
what implies that the notion of gravitational energy density makes no 
sense. Yet in a field theory in Minkowski space the field energy density is 
equal to the Hamiltonian density and the latter is (for known fields) 
positive definite. In the canonical ADM formalism in general relativity the 
canonical momenta are defined in an intricate way (including constraints), 
not akin to that in point mechanics and the total ADM energy is to large 
extent independent of the detailed form of the Hamiltonian density (which 
is indefinite). Therefore in general relativity the relationship between 
stability (understood as the positivity of energy, see below) and the form 
of the Hamiltonian density is very indirect, practically broken. In metric 
NLG theories the Legendre transformations from Jordan frame to 
Helmholtz--Jordan frame (HJF) and Einstein frame (EF)  
map the higher derivative theory to Einstein gravity plus nongeometric 
components of the multiplet which dynamically act as some matter fields, 
therefore the stability problem in these theories is reduced to that in the 
latter theory. The Ostrogradski theorem may rather serve as a warning that 
some troubles may appear there and in fact troubles were found (the 
ghost--like behaviour of the massive spin--2 component of gravity) without 
resorting to it. Note that the notion of "inherently unstable theory" is 
imprecise: stability always concerns a given solution. And what is really 
required from a viable gravity theory is existence of a stable ground state 
solution; stability of excited states is a different problem. \\

In the physical literature there is some confusion concerning stability 
since there are actually two notions of stability: dynamical stability 
(stability of evolution) meaning that there are no growing modes and 
stability as a consequence of positivity of total energy. It has been 
believed for a long time that the two notions are identical and since 
investigations of energy are relatively easier the research was first 
centered on it. Stability in the context of energy was developed in a 
series of papers which will be referred here to as "classical works". 
Positivity of the ADM energy implies stability of Minkowski space. 
The notion of this energy was then extended to the Abbott--Deser (AD) 
energy for the spacetimes which are asymptotically de Sitter or 
anti--de Sitter \cite{AbD}. Applying this notion it was shown that 
vacuum dS is linearly stable \cite{AbD} while AdS space is both 
linearly and nonlinearly stable in vacuum \cite{AbD} and in presence 
of any matter satisfying the dominant energy condition (DEC) in any 
dimension $d\geq 4$ \cite{GHW,HM}. However it was found that 
stability does not necessarily result from the positivity of energy:  
there are situations in which the positive energy theorem 
holds and instabilities develop \cite{GP}. Thus dynamical stability 
(no growing in time perturbation modes) and positivity of 
energy are quite different unrelated things. Stability of evolution 
requires mathematically rigorous investigations.\\

In the rigorous approach it was shown that Minkowski space is globally 
dynamically stable: in vacuum \cite{ChK}, in presence of the 
electromagnetic field \cite{Zi} or of the linear massless 
scalar field \cite{LR}. Vacuum de Sitter space is globally stable in 
four \cite{Fr1} and any larger even number of dimensions \cite{An1}. 
Inclusion of matter is difficult: global stability of dS space was 
proved only in the case of Yang--Mills fields (in $d=4$) \cite{Fr2} 
and for a scalar field with a very specific potential \cite{HR}; its 
stability for all other forms of matter is unknown. 
Even less is rigorously known about stability of anti--de Sitter space:  
it is globally linearization stable for the Maxwell and linear scalar 
field \cite{IW} and for the vacuum case Friedrich \cite{Fr4} proved 
finite time nonlinear stability. There are no rigorous global results, 
it is only believed that vacuum AdS space is dynamically stable and 
nothing has been investigated in the case of self--interacting 
scalar fields. \\

While the fully reliable rigorous results are quite modest from the 
standpoint of a physicist dealing with gravitational fields generated by 
a rich variety of matter sources, the classical theorems based on the 
positivity of energy are, from the viewpoint of mathematicians, of rather 
little reliability \cite{An3}. In proving the rigorous theorems only the 
exact field equations are relevant and the dominant energy condition does 
not play explicitly any role. However in the few cases where matter sources 
are present, DEC does hold. It is therefore reasonable to conjecture that 
Minkowski, de Sitter and anti--de Sitter spaces are globally nonlinearly 
stable only if any self--gravitating matter does satisfy the condition. 
The conjecture is supported by outcomes found in the linear approximation 
to semiclassical general relativity where the expectation 
value $<0|T_{\mu\nu}|0>$ cannot satisfy DEC due to the particle creation 
by the gravitational field. In the presence of the electromagnetic, 
neutrino and massless scalar fields Minkowski space is linearly unstable 
\cite{Ho} and similarly a minimally coupled quantum scalar field 
renders Sitter space linearly unstable \cite{TH}.\\ 

All the aforementioned papers deal with solutions to Einstein field 
equations. Recently Faraoni \cite{Fa1} studied stability of vacuum dS 
space in restricted NLG theories in Jordan frame for the fourth--order 
field equations. The dS metric can be presented in the form of the 
spatially flat Robertson--Walker spacetime and he has applied the gauge 
invariant formalism of Bardeen--Ellis--Bruni--Hwang for perturbations of 
Friedmann cosmology. The formalism works for any field equations in this 
background and he proves linearization stability of dS space: scalar and 
tensor metric perturbations are fading or oscillating at late times 
provided the Lagrangian $L=f(R)$ satisfies some inequality. In this 
formalism the physical meaning of this crucial inequality is unclear. 
It turns out that the condition is equivalent to the condition that the 
(positive) potential for the scalar component of gravity in Einstein frame 
attains minimum at dS space being a ground state solution, see sect. 6. 
The BEBH formalism does not apply to perturbations of AdS space since its 
metric cannot be expressed as the spatially flat R--W spacetime. It is 
interesting to see that in most papers on NLG theories it is assumed that 
a curved ground state is necessarily dS space while AdS space is omitted 
without mention\footnote{AdS space is mentioned as a possible ground 
state e.g.~in \cite{Di, NO1}.}. \\

We shall investigate stability of the maximally symmetric ground state  
solutions in various NLG 
theories in a coordinate independent manner. We presume that the classical 
works provide the correct assumptions under which the dynamical stability 
of these solutions will be rigorously proved in future. We shall work in 
Einstein frame where the only source for the metric is the scalar field 
component of gravity since on physical grounds it is stability of pure 
gravity that is crucial. Moreover we argue in section 4 that inclusion of 
matter (e.g.~perfect fluid) does not affect stability of the solutions. 
We emphasize that stability of a candidate ground state 
solution is independent of which frame is regarded as physical since 
boundedness of solutions remains unaltered under Legendre transformations. 
The method based on positivity of total ADM or AD energy works directly 
only in Einstein frame. 
The energy--momentum tensor of the scalar satisfies the 
dominant energy condition if and only if its potential is nonnegative. Thus 
satisfying DEC for the field becomes an effective viability criterion for 
restricted NLG theories. 

\section{Candidate ground state solutions}
We shall now investigate existence of candidate ground state (CGS) solutions, 
i.e., maximally symmetric (dS, AdS or Minkowski space) solutions in a 
restricted NLG theory with $L=f(R)$ for arbitrary $f$. A CGS solution 
becomes a true physical ground state solution (named vacuum) if it is stable. 
We assume that the Lagrangian has the same dimension as the curvature 
scalar, $[f(R)] =[R] = (\textrm{length})^{-2}$, and the signature is 
$(-+++)$. The field equations in Jordan frame are 
\begin{equation}
E_{\mu\nu}(g) \equiv f'(R)R_{\mu\nu} - \frac{1}{2}f(R)g_{\mu\nu}-
\nabla_{\mu}\nabla_{\nu}f'(R) +g_{\mu\nu}\Box f'(R) =0 
\end{equation}
or 
\begin{displaymath}
f'(R)R_{\mu\nu} - \nabla_{\mu}\nabla_{\nu}f'(R) 
+\frac{1}{3}g_{\mu\nu}\left[\frac{1}{2}f(R) -R f'(R)\right] =0, 
\end{displaymath}
here $f'\equiv \frac{df}{dR}$ and $\Box\equiv g^{\mu\nu}
\nabla_{\mu}\nabla_{\nu}$. In general $f(R)$ 
cannot be everywhere smooth and the nonlinear equations (1) require $f$ 
be piecewise of $C^3$ class\footnote{Any Lagrangian is determined up to 
a divergence of a vector field made up of the dynamical variables. If the 
gravitational Lagrangian is to be a scalar function of the Riemann tensor 
invariants alone and involve no derivatives of the curvature, the 
Lagrangian is determined up to a constant multiplicative factor. The 
factor must be fixed if any matter is minimally coupled to gravity in 
JF.}.\\

A CGS solution exists if and only if the field equations (1) admit a 
class of Einstein spaces, $R_{\mu\nu}=\frac{1}{4}\lambda g_{\mu\nu}$, 
for some curvature scalar $R=\lambda$, as solutions. Since $\lambda=
\textrm{const}$ and assuming that $\lambda$ lies in the interval where 
$f(\lambda)$, $f'(\lambda)$, $f''(\lambda)$ and $f'''(\lambda)$ are 
finite, eqs. (1) reduce to an algebraic equation 
\begin{equation}
\lambda f'(\lambda) - 2 f(\lambda) =0.
\end{equation}
This equation was first found by Barrow and Ottewill \cite{BO} and then 
rediscovered many times. In general this equation has many solutions and to 
each solution $\lambda = \lambda_i$ there corresponds a whole class of 
Einstein spaces containing a maximally symmetric spacetime, being dS for 
$\lambda_i >0$, AdS for $\lambda_i <0$ or Minkowski space ($\cal{M}$) 
for $\lambda_i =0$. For some $\lambda_i$ the maximally symmetric space may 
be stable. Each stable ground state (vacuum) defines a separate dynamical 
sector of the theory. Multiplicity of vacua for a $L=f(R)$ gravity was 
first noticed in \cite{HOW1}. \\

We view (2) as an algebraic equation and assume that it has at most the 
countable number of solutions. Not every function $f(R)$ admits a solution 
to (2). First note the degenerate case where any value of $\lambda$ is a 
solution (uncountable number of solutions); this occurs when (2) is 
viewed as a differential equation for $f$. Then $f(R)= a R^2$ for any 
constant $a\ne 0$ \cite{BO}. In the following we make some comments 
on this degenerate case. The eq. (2) has no solutions if its LHS defines 
a function of $R$, $Rf'(R) - 2f(R)\equiv F(R)$, which nowhere vanishes. 
Treating this definition as a differential equation for $f$ (for a given 
$F$) one finds that any Lagrangian which admits no CGS solutions is of 
the form 
\begin{equation}
f(R) = a R^2 \pm R^2 \int\frac{F(R)}{R^3} \ud R
\end{equation}
with arbitrary $F(R)>0$ everywhere. Examples.\\
1. Let $F(R)= C e^{bR}$ with $b,C\ne 0$. Then 
\begin{displaymath}
f(R) = aR^2 + C\left[-\frac{1}{2}e^{bR}(1+bR) + \frac{b^2}{2}R^2 
\int\frac{e^{bR}}{R} \ud R\right],
\end{displaymath}
the latter integral is non--elementary.\\
2. For 
\begin{displaymath}
F(R) = c_0 + \sum_{n=1}^{\infty} c_{2n}R^{2n}
\end{displaymath}
with $c_0$ and $c_{2n}>0$ one gets 
\begin{displaymath}
f(R) = -\frac{c_0}{2}+ aR^2 + c_2 R^2\ln |R| +\frac{1}{2}
\sum_{n=2}^{\infty}\frac{c_{2n}}{n-1}R^{2n}. 
\end{displaymath}
Any gravity theory with a Lagrangian of the form (3) is unphysical and 
should be rejected. Clearly there are infinitely many functions $f(R)$ 
admitting solutions to (2) and thus possibly possessing a stable 
vacuum. A few examples of these Lagrangians.\\

1. $f(R)= R+ aR^2 + \alpha^{-2}R^3$, $\alpha>0$. There are three CGS 
solutions: $\cal{M}$ ($\lambda_1 =0$), dS ($\lambda_2 = +\alpha$) and 
AdS ($\lambda_3 = -\alpha$). \\
2. $f(R)= -2\Lambda + R + aR^2 + bR^3$. \\
Eq. (2) becomes a cubic equation $\lambda^3 -\lambda/b + 4\Lambda/b =0$ 
having 3 distinct real roots if $4\Lambda^2 <(27b)^{-1}$ and one real 
root $\lambda_1$ (and $\lambda_2= \bar{\lambda}_3$ complex) if 
$4\Lambda^2 >(27b)^{-1}$. In the limiting case $4\Lambda^2 =(27b)^{-1}$ 
there are 3 real roots with $\lambda_1=\lambda_2$ and $\lambda_3 \ne 
\lambda_1$. The solutions are independent of the coefficient $a$. In the 
case $b=0$ there is a unique CGS solution $\lambda=4\Lambda$, being dS 
or AdS, the same as in Einstein theory. \\
3. For $f(R) = \frac{1}{a}e^{aR}$, $a>0$, there is a dS space with 
$\lambda = 2/a$ as a unique CGS solution. \\
4. For $f(R) =a\sqrt{R} +b R^{7/3}$, $a, b>0$, there are two CGS 
solutions: $\lambda_1=0$ and $\lambda_2= +(\frac{9a}{2b})^{6/11}$. \\

The last example illustrates a general rule: if $f(0)=0$ then 
$\lambda=0$ is always a solution of (2), i.e., $\cal{M}$ is a CGS 
solution even if $f'(0)$ is divergent. In fact, if $f'(0)\to \pm
\infty$ the term $Rf'(R)$ may a priori either vanish at $R=0$ or 
diverge either logarithmically or as inverse of a power law. If 
$Rf'(R)=R^{-n}$ for some $n>0$ then $f= c -\frac{1}{n}R^{-n}$, 
while for $Rf'(R)=(\ln R)^n$, $n>0$, one finds $f=c +\frac{1}{n+1}
(\ln R)^{n+1}$; in both cases $f(0)$ is divergent. For $f(0)=0$ and 
$f'(0)\to \pm\infty$ the leading term in $f$ near $R=0$ is $R^a$, 
$0<a<1$, and then $Rf'(R)\to 0$. However if $f'(0)$ and/or $f''(0)$ 
is divergent the method for establishing whether $\cal{M}$ is stable 
does not work. \\

We emphasize that in order to investigate the dynamics of a restricted 
NLG theory one needs \emph{exact\/} solutions of eq. (2). We shall see 
that stability of a CGS solution is determined by the values of 
$f'(\lambda)$ and $f''(\lambda)$. In principle to check stability it is 
sufficient to find numerically an approximate solution $\lambda$ to eq. 
(2) and then approximate values of $f'(\lambda)$ and $f''(\lambda)$. 
Also the mass of the scalar component of gravity is determined by these 
two numbers. However an exact solution is necessary to calculate the 
scalar field potential both in Helmholtz--Jordan and Einstein frames; 
otherwise one gets only approximate equations of motion in these 
frames as is shown in the following example:\\
$L=f(R)= a\sin \frac{a}{R}$, $a>0$. Introducing a dimensionless quantity 
$x\equiv \frac{a}{\lambda}$ one finds that eq. (2) reads $x\cos x+2
\sin x=0$. $\cos x=0$ is not a solution and the equation may be written 
as $x+2\textrm{tg}\,x=0$. The obvious root is $x=0$, but it corresponds to 
$R=\lambda=\infty$ and this solution must be rejected on physical grounds. 
In the interval $-\pi/2<x<\pi/2$ where tangens is continuous the functions 
$x$ and $\textrm{tg}\,x$ are of the same sign and the equation has no 
solutions. In each interval $(n-1/2)\pi<x<(n+1/2)\pi$, $n=\pm 1, \pm 2,
\ldots$, the equation has exactly one solution which may be determined 
numerically. The scalar component of gravity is defined as 
$p=df/dR$ and to determine the potential for $p$ one needs to invert 
this relation to get $R=r(p)$. In the present example $p=-(a/R)^2 
\cos a/R$ and though this relation is in principle invertible (since 
$f''(R)\ne 0$ and $f''$ vanishes only at separate points where 
$\textrm{tg}\,a/R=2R/a$), it cannot be inverted analytically in any of 
the intervals. One sees that exact solvability of eq. (2) is often 
correlated to exact invertibility of the definition $p=f'(R)$. We 
conclude that the condition of exact analytic solvability of eq. (2) 
is of crucial importance and in practice imposes stringent restrictions 
on the Lagrangians excluding many simple combinations of elementary 
functions. A further constraint will be imposed in the next section. \\

Finally we make two remarks on the field equations (1).\\
Firstly, recall that for cosmologists the most attractive Lagrangians 
are those containing inverse powers of $R$ rather than being 
polynomials in $R$. In consequence the coefficients of fourth order 
derivatives in (1) are rational functions and this implies that one 
should deal with great care with various terms in these equations in 
order to avoid multiplying or dividing by zero\footnote{We stress that 
this is not trivial. In a frequently quoted paper \cite{DK} the trace 
of eqs. (1) for a Lagrangian $R-1/R$ was multiplied by $R^3$ giving 
rise to a scalar equation for $R$ admitting $R=0$ as a solution and 
thus Minkowski space; further considerations of the work were based 
on perturbations of this spacetime. Actually the field equations for 
this Lagrangian have only dS and AdS spaces as CGS solutions. This 
"curvature instability" found in \cite{DK} has been generalized to 
many other functions $f(R)$ without checking if Minkowski space is 
a solution and is even regarded as advantage of the Palatini formalism 
over the purely metric gravity theories \cite{Sot}. This 
error of introducing or omitting some classes of solutions by 
multiplying the field equations by a power of $R$ may be traced back 
to Bicknell \cite{Bi}.}. For simplicity we demonstrate it on a toy 
model. Suppose that field equations read 
\begin{equation}
R_{\mu\nu}+\frac{1}{R^2}\Box R_{\mu\nu}=0.
\end{equation}
Multiplying them by $R^2$ one gets
\begin{equation}
R^2 R_{\mu\nu}+\Box R_{\mu\nu}=0
\end{equation}
and a class of solutions to these equations is given by $R_{\mu\nu}=
\psi_{\mu\nu}\ne 0$ where the tensor is traceless, $R=\psi\equiv 
g^{\mu\nu}\psi_{\mu\nu}=0$ and satisfies $\Box\psi_{\mu\nu}=0$. 
However $\psi_{\mu\nu}$ is not a solution to (4) since the LHS of 
these equations is then $\psi_{\mu\nu} + 0/0$. A class of solutions 
to (4) is of the form $R_{\mu\nu}=\phi_{\mu\nu}\ne 0$ and 
$\Box\phi_{\mu\nu}=-\phi^2\phi_{\mu\nu}$ with $\phi\equiv
g^{\mu\nu}\phi_{\mu\nu}$; clearly these are also solutions to (5). 
Furthermore, any spacetime satisfying $R_{\mu\nu}=0$ is a solution  
to both (4) and (5). At first sight this is not since
the second term in (4) becomes divergent. One may however give a 
precise meaning to this term by trying an Einstein 
space, $R_{\mu\nu}=(\lambda/4)g_{\mu\nu}$, then $\Box R_{\mu\nu}
\equiv 0$ and the eqs. (4) reduce to $\lambda g_{\mu\nu}=0$ so that 
$R_{\mu\nu}=0$ actually are solutions. In conclusion, by replacing 
the correct equations (4) by allegedly equivalent equations (5) 
one introduces a class of false solutions $R_{\mu\nu}=
\psi_{\mu\nu}$. \\

Secondly, we comment on the cosmological constant \cite{CB2}. In 
metric NLG theories this notion has a rather limited sense. In general 
relativity $\Lambda$ is both the constant appearing in the 
Einstein--Hilbert Lagrangian, $\Lambda = -\frac{1}{2}L(0)$, and the 
curvature of the unique maximally symmetric ground state, $\Lambda =
R/4$. If $f(0)\ne 0$ is finite in an NLG theory one may define $\Lambda$ 
as $-\frac{1}{2}f(0)$, however there is at least one CGS 
solution with the curvature $R=\lambda\ne 0$ whose value is 
independent of the value $f(0)$ (in the sense that the function $F(R)
\equiv Rf'(R) -2f(R)$ may be freely varied near $R=0$  provided 
$F(0)\ne 0$ is preserved, then $R=\lambda$ remains the solution of 
(2)). Alternatively, $\Lambda$ may be defined as $\lambda/4$ for each 
vacuum (stable ground state), then $\Lambda$ has different values in 
different sectors of the theory. However this cosmological constant is 
related solely to the vacuum and does \emph{not\/} appear as a 
parameter in other solutions to the field equations (1). We 
therefore shall not use this notion. 

\section{The field equations including matter}
We shall now express the field equations in the form appropriate for 
investigating stability of the CGS solutions. Detailed calculations 
based on the general formalism \cite{MFF1, JK} are given in \cite{MS1}. 
The scalar component of the gravitational doublet is defined in HJF as 
$p\equiv \frac{df}{dR}$, this canonical momentum is dimensionless. The 
definition is inverted to give the curvature scalar $R$ as a function 
of $p$, $R(g)=r(p)$, i.e., 
\begin{displaymath}
f'(R)\big|_{R=r(p)} \equiv p.
\end{displaymath}
The inverse function $r(p)$ exists iff $f''(R)\ne 0$. The pure gravity 
Helmholtz action 
\begin{displaymath}
S_{HJ} =\int\ud^4 x\,\sqrt{-g}\, L_H(g, p)
\end{displaymath}
with $L_H= p[R(g)-r(p)] +f(r(p))$ (see Paper I) gives rise to the 
field equations 
\begin{equation}
G_{\mu\nu}(g) =\theta_{\mu\nu}(p,g) \equiv \frac{1}{p} 
\nabla_{\mu}\nabla_{\nu}p -
\frac{1}{6}\left[\frac{1}{p}f(r(p)) +r(p)\right] g_{\mu\nu}
\end{equation}
and 
\begin{equation}
\Box p -\frac{2}{3}f(r(p))+ \frac{1}{3}p r(p)=0.
\end{equation}
By taking trace of (6) and employing (7) one recovers the 
relation $R(g)=r(p)$. The effective energy--momentum tensor for $p$ 
contains a linear term signalling that the energy density is 
indefinite and deceptively suggesting that all solutions, including 
the CGS ones, are unstable \cite{Woo}. However $\theta_{\mu\nu}$ 
turns out unreliable in this respect and to study stability one makes 
the transformation from HJF to Einstein frame being a mere change of 
the dynamical variables. It consists of a conformal map of the 
metric, 
\begin{displaymath}
g_{\mu\nu} \to \tilde{g}_{\mu\nu}\equiv p g_{\mu\nu},
\end{displaymath}
and a redefinition of the scalar, 
\begin{displaymath}
p\equiv \exp\left(\sqrt{\frac{2}{3}}\kappa \phi\right)
\end{displaymath}
or $\phi =\sqrt{\frac{3}{2}}\frac{1}{\kappa}\ln p$, with $\kappa$ 
being a dimensional constant to be specified later\footnote{In Paper I, 
for simplicity we have put $\kappa =1$ in eqs. (8), (9) of that paper 
and the 
definition of $\phi$.}. Under the transformation of the variables the 
action integrals in HJF and EF are equal, 
\begin{equation}
S_{HJ} = S_E= \int\ud^4 x\,\sqrt{-\tilde{g}}\, \tilde{L}_H(\tilde{g}, 
p(\phi)),
\end{equation}
what defines $\tilde{L}_H$. 
To get the total Lagrangian precisely as in general relativity one 
introduces an equivalent Lagrangian proportional to $\tilde{L}_H$,
\begin{equation}
L_E \equiv \frac{1}{2\kappa^2 c}\tilde{L}_H =
\frac{1}{2\kappa^2 c} \tilde{R}(\tilde{g}) + \frac{1}{c} L_{\phi}
\end{equation}
and sets $(2\kappa^2 c)^{-1}\equiv c^3/(16\pi G)$ or $\kappa^2 =8\pi G/
c^4$. 
Hence $\phi$ is a minimally coupled scalar field with a 
self--interaction potential,
\begin{equation}
L_{\phi} = - \frac{1}{2}\tilde{g}^{\alpha\beta}
\phi_{,\alpha}\phi_{\beta} -\frac{1}{2\kappa^2}\left[\frac{r(p)}{p} -
\frac{f(r(p))}{p^2}\right] \equiv
- \frac{1}{2}\tilde{g}^{\alpha\beta}
\phi_{,\alpha}\phi_{\beta} - V(p(\phi)).
\end{equation}
The constant $\kappa$ determines the dimension of $\phi$, $[\phi]=
\textrm{g}^{1/2}\textrm{cm}^{1/2}\textrm{s}^{-1}$, while $V$ acquires 
dimensionality of energy density. The field equations following from 
(9) are 
\begin{equation}
\tilde{G}_{\mu\nu}(\tilde{g}) = \kappa^2 T_{\mu\nu}(\phi, \tilde{g}) 
= \kappa^2(\phi_{,\mu}\phi_{,\nu} - \frac{1}{2}\tilde{g}_{\mu\nu}
\tilde{g}^{\alpha\beta}\phi_{,\alpha}\phi_{,\beta} -
\tilde{g}_{\mu\nu} V(\phi))
\end{equation}
and
\begin{equation}
\stackrel{\sim}{\Box}\!\!\phi = \frac{\ud V}{\ud \phi} =
\sqrt{\frac{2}{3}}\kappa p \frac{\ud V}{\ud p}.
\end{equation}
Solutions for a self--interacting scalar field in general relativity 
were studied in many papers, however they are not solutions to eqs. 
(10)--(12) since the potential (10) is in most cases different from the 
potentials appearing in those papers. For example, an exponential 
potential $V_0 \exp(-\alpha\kappa\phi)$ with constant $\alpha$ was 
investigated in a number of works (see e.g.~\cite{HeW}); in terms of 
the scalar $p$ it reads $V_0 p^{-\alpha}$, but there are no simple 
Lagrangians $L=f(R)$ generating this potential via eq. (10). 
Recall that as long as one considers pure gravity, i.e. there is no 
minimally coupled matter in JF, the original Lagrangian $L= f(R)$ is 
determined up to an arbitrary constant factor A. Let $\bar{f}(R)
\equiv Af(R)$. Then $\bar{p}\equiv \bar{f}'(R) =Ap$ and the inverse 
relation is $R(g)=\bar{r}(\bar{p})$. On the other hand $R(g)=r(p)$ 
so that $\bar{r}(\bar{p})=r(p)= r(\bar{p}/A)$. This implies $L_H
(g, \bar{p})= AL_H(g,p)$, the conformal factor $\bar{p}$ generates 
in EF the metric $\hat{g}_{\mu\nu}=A\tilde{g}_{\mu\nu}$ and 
$\tilde{L}_H(\hat{g},\bar{p})= A^{-1}\tilde{L}_H(\tilde{g}, p)$.\\ 
The conformal map should not alter the signature of the metric, thus 
one requires $p>0$. In general $f'(R)$ cannot be positive for all 
$R$ and it is sufficient to require that the map preserve the 
signature at the CGS solutions, i.e., $p(\lambda)=f'(\lambda)>0$ 
for each solution of eq. (2). Then $p>0$ in some neighbourhood of 
$R=\lambda$. If $p(\lambda)<0$ one should take the Lagrangian $L=
-f(R)$. It may occur for some $f(R)$ having multiple solutions of 
(2) that $p(\lambda_i)>0$ and $p(\lambda_j)<0$ for $i\ne j$, then 
one should appropriately choose the sign of $L$ at each sector of 
the theory separately. We shall assume that this has been 
done\footnote{One may try a simplification by choosing $L(R)=
\frac{f(R)}{f'(R)}$, then $L'(\lambda)=1$. Actually this choice 
does not simplify the expressions for derivatives of the potential 
$V$ and we shall not apply it.} and $p(\lambda_i)=f'(\lambda_i)>0$.\\

The transformation from HJF to Einstein frame exists in a 
neighbourhood of a CGS solution with $R=\lambda$ iff $f'(\lambda)
\ne 0$. If $f'(\lambda)=0$ the EF does not exist and the method of 
checking stability of the CGS solution does not apply. From 
$\lambda f'(\lambda) - 2f(\lambda)=0$ it follows that $f(\lambda)
=0$ and assuming that $f$ is analytic around $R=\lambda$ it has a 
general form 
\begin{equation}
f(R)=\sum_{n=2}^{\infty} a_n (R-\lambda)^n 
\end{equation}
for any real $\lambda$. Notice that the degenerate Lagrangian $L=R^2$ 
belongs to this class. This class of singular Lagrangians needs a 
separate treatment (see section 6) and we assume that $f(R)$ is not 
of the form (13). \\
For Lagrangians which are different from (13) the potential $V(\phi)$ 
in EF is not a constant. To prove it one assumes that $V=
\textrm{const}$ and determines the corresponding $f(R)$. From (10) 
one gets 
\begin{equation}
r(p)= Cp +\frac{f(r)}{p}
\end{equation}
where $C\equiv 2\kappa^2 V$ and one differentiates this equation with 
respect to $f$ employing 
\begin{displaymath}
\frac{\ud r}{\ud f}=\left(\frac{\ud f}{\ud r}\right)^{-1}= \frac{1}
{p}. 
\end{displaymath}
One finds 
\begin{displaymath}
\frac{\ud r}{\ud f}=\frac{1}{p}= C\frac{\ud p}{\ud f} +\frac{1}{p}-
\frac{f}{p^2}\frac{\ud p}{\ud f} \quad \textrm{or} \quad 
\frac{\ud p}{\ud f}(C-\frac{f}{p^2})=0.
\end{displaymath}
\begin{displaymath}
\textrm{Since} \qquad 
\frac{\ud p}{\ud f}=\left(\frac{\ud f}{\ud r} 
\frac{\ud r}{\ud p}\right)^{-1} \ne 0 \qquad \textrm{this yields}
\end{displaymath}
\begin{displaymath}
f(r(p))= C p^2.
\end{displaymath}
Inserting this value of $f$ into (14) yields $r(p)= 2Cp$ and 
substituting $p=\frac{r}{2C}$ from the latter relation back to $f=
Cp^2$ one finally finds $f=\frac{r^2}{4C}$. Using $R(g)=r(p)$ one 
arrives at $f(R)=\frac{R^2}{4C}$ for any real $C\ne 0$, i.e., the 
degenerate Lagrangian. In particular the potential cannot vanish 
identically. In fact, $V=0$ implies $r(p)= f(r)/p$. Differentiating 
this relation with respect to $r$ under assumption that $f'(r)\ne 0$ 
and $f''(r)\ne 0$ (the condition for $r(p)$ to exist) one arrives at 
$ff''/p^2=0$ implying $f''=0$. This contradiction shows that $V\not
\equiv 0$.\\
For admissible Lagrangians the potential is variable and this feature 
will be used to establish stability. \\

Finally we comment on stability of a CGS solution in the presence of 
some matter. In our opinion it is the stability of pure gravity (only 
the metric and the scalar) that is crucial for physical viability of 
the theory while exotic forms of matter violating DEC can make the 
ground state unstable even in general relativity as it found in the 
two examples mentioned in section 2. Yet recently there appeared 
claims (see e.g.~\cite{Sei}) that the very presence matter (perfect 
fluid stars) renders $f(R)$ gravity unstable. We show now that this 
is not the case. The point is that the property of the variational 
matter energy--momentum tensor (the stress tensor for short) to satisfy 
DEC is preserved under a conformal map of the metric. If one assumes 
that Jordan frame is physical and minimally couples a given species 
of matter $\Psi$ in this frame, then the stress tensor $t_{\mu\nu}
(\Psi, g)$satisfies DEC by assumption. The field equations (1) for 
the metric read then\footnote{The choice of a constant coefficient in 
front of $f(R)$ in the Lagrangian should be determined by a Newtonian 
limit of the theory. This can be unambiguously done in the case where 
Minkowski space is the ground state. If the CGS solution under 
consideration is dS or AdS, the Newtonian limit is not well defined 
and the coefficient is undetermined. This trouble does not affect 
the present argument.}
\begin{equation}
f'(R)R_{\mu\nu} - \frac{1}{2}f(R)g_{\mu\nu}-
\nabla_{\mu}\nabla_{\nu}f'(R) +g_{\mu\nu}\Box f'(R) =
\kappa^2 t_{\mu\nu}(\Psi, g). 
\end{equation}
The conformal map $\tilde{g}_{\mu\nu}= p g_{\mu\nu}$ makes the 
matter Lagrangian explicitly dependent on the scalar gravity $p$ and 
the stress tensor in EF for $\Psi$ alone cannot be unambiguously 
derived from it. It is therefore convenient to express the 
gravitational field equations in both the frames in terms of 
$t_{\mu\nu}$ which is already defined as the variational one in 
terms of the physical metric (i.e.~in JF). The metric field 
equations in EF replacing (11) are then \cite{MS1}
\begin{equation}
\tilde{G}_{\mu\nu}(\tilde{g}) = \kappa^2 T_{\mu\nu}(\phi, \tilde{g}) 
+ \frac{\kappa^2}{p} t_{\mu\nu}(\Psi, \frac{1}{p}\tilde{g}).
\end{equation}
Since $p>0$ in a vicinity of the CGS solution and DEC holds for both 
the stress tensors in EF\footnote{This frame invariance of DEC does 
not apply to the scalar gravity $p$ since in any frame its stress 
tensor is defined as that equal in pure gravity (no matter) to the 
Einstein tensor of the given metric.}, it also holds for the total 
stress tensor $T_{\mu\nu} +\frac{1}{p}t_{\mu\nu}$. This means that 
matter cannot destroy stability of the ground state if it is stable 
in pure gravity theory. We comment on the instability found in 
\cite{Sei} in section 7.\\

\section{Positive energy theorem for anti--de Sitter space}
We emphasize that the applied here method of proving stability of dS, 
AdS or $\cal{M}$ spaces is based on the assumption that the scalar 
component of gravity satisfies in EF the dominant energy condition, 
what is equivalent to $V(\phi)\geq 0$. The fact that it implies 
positivity of total ADM or AD energy is not used. Nevertheless we shall 
consider for the moment this energy. In \cite{MS1} we proved that if 
$L=f(R)$ is analytic at $R=0$ and its expansion is $L=R +aR^2+\ldots$ 
and the potential $V(\phi)$ in EF is non--negative, the ADM energy of a 
spacetime which is asymptotically flat is the same in both Jordan and 
Einstein frames and is non--negative. Near $\cal{M}$ the potential 
behaves as $V=\frac{1}{2}aR^2 +O(R^3)$, whence $V>0$ for $a>0$. An 
analogous positive--energy theorem may be proved in restricted NLG 
theories for spacetimes which are asymptotically AdS space. The case of 
spacetimes which asymptotically converge to de Sitter space is more 
complicated because dS is not globally static and we 
disregard it.\\

Let $\bar{g}_{\mu\nu}$ be the metric of AdS space in the following 
coordinates:
\begin{equation}
\ud \bar{s}^2= \bar{g}_{\mu\nu}\ud x^{\mu} \ud x^{\nu} =-(1+\frac{r^2}
{a^2})\,\ud t^2 +(1+\frac{r^2}{a^2})^{-1}\ud r^2 +r^2(\ud \theta^2+
\sin^2\theta\, \ud \varphi^2),
\end{equation}
the cosmological constant is $\Lambda=-\frac{3}{a^2}$, $a=\textrm{const}
>0$ and $\bar{R}=\lambda= 4\Lambda$. Let $g_{\mu\nu}$ be a solution of 
the field equations (1) in JF which asymptotically approaches AdS metric 
(17), $g_{\mu\nu}=\bar{g}_{\mu\nu}+ h_{\mu\nu}$. Clearly $g_{\mu\nu}$ is 
a solution to Einstein field equations $G_{\mu\nu}(g)=\theta_{\mu\nu}$ 
in HJF, then the Abbott--Deser approach \cite{AbD} applies and the 
total energy of the fields $g_{\mu\nu}$ and $p$ is given by their formula, 
which in the case of (17) reduces to 
\begin{eqnarray}\label{n18}
E_{AD}[g] & = & \frac{c^4}{16\pi G}\lim_{r \to\infty}\int\sin\theta 
\ud \theta\,\ud \phi [-r^2\partial_1 h_{00} +\frac{r^6}{a^4}
\partial_1 h_{11}+ \frac{r^2}{a^2}(\partial_2 h_{12}+
 \frac{1}{\sin^2\theta}\partial_3 h_{13})
\nonumber \\
& &+ 3r h_{00} +\frac{3}{a^4}r^5 h_{11}-\frac{r}{a^2}
(h_{22} +\frac{h_{33}}{\sin^2\theta}) +\frac{r^2}{a^2}h_{12}\,
\textrm{ctg}\,\theta],
\end{eqnarray}
here $x^i=(r, \theta, \phi)$ and the timelike Killing vector in the 
Abbott--Deser formula is chosen as $\xi^{\mu}=\delta^{\mu}_0$, then 
its normalization at $r=0$ is $\xi^{\mu}\xi_{\mu}=-1$. In general all 
the components of $h_{\mu\nu}$ are algebraically independent and the 
requirement that separately each term in the integrand of (18) gives 
rise to a finite integral (what amounts to requiring that each term be 
independent of $r$) provides the asymptotic behaviour of:\\
$h_{00}$, $h_{22}$ and $h_{33}$ are of order $r^{-1}$, $h_{11}=
O(r^{-5})$ and $h_{12}=O(r^{-2})=h_{13}$.\\ A spacetime being asymptotically 
anti--de Sitter space is defined in \cite{BGH} and according to this 
definition a solution approaches AdS slower than is required by 
finiteness of its energy. We assume that the six components of $h_{\mu
\nu}$ behave as shown above while the remaining four components, which 
do not enter the energy integral, tend to AdS as in the definition in 
\cite{BGH}, $h_{01}=O(r^{-1})$ and $h_{02}$, $h_{03}$ and $h_{23}$ are 
$O(r)$. Under these assumptions the scalar $R(g)$ for a solution with 
finite energy approaches $\bar{R}=4\Lambda$ as $R\to 4\Lambda +
O(r^{-2})$. \\

In Einstein frame an analogous integral expression for $E_{AD}[\tilde
{g}]$ holds for the corresponding solution $\tilde{g}_{\mu\nu}$ with 
$h_{\mu\nu}$ replaced by $\tilde{h}_{\mu\nu}=p(R) h_{\mu\nu}$. For 
$r \to\infty$ the conformal factor is $p=f'(R)=f'(\bar{R}+O(r^{-2}))
=f'(4\Lambda) +O(r^{-2})$ (assuming that $f''(4\Lambda)\ne 0$ and 
finite), whence $E_{AD}[\tilde{g}]= f'(4\Lambda)E_{AD}[g]$ is 
finite. This energy is positive according to the positive energy 
theorem in general relativity provided $V(\phi)>0$. Since $f'(4
\Lambda)>0$ by assumption, we get that in spite of the indefiniteness 
of the tensor $\theta_{\mu\nu}(g, p)$ in HJF the 
\emph{positive--energy theorem for restricted NLG theories\/} holds:\\
(i) if $L=f(R)$ admits AdS space with $\bar{R}=4\Lambda<0$ as a 
solution, (ii) $f'(4\Lambda)>0$ and $f''(4\Lambda)\ne 0$ is finite, 
(iii) the potential $V(\phi)$ in EF is non--negative and 
(iv) a solution $g_{\mu\nu}$ in JF or equivalently the pair 
$(g_{\mu\nu}, p)$ in HJF tends sufficiently quickly to AdS space for 
$r\to\infty$, then the total energy in JF is equal to the AD energy 
in HJF and positive and proportional to that in EF, \\
\begin{displaymath}
E_{AD}[g]=\left(f'(4\Lambda)\right)^{-1} E_{AD}[\tilde{g}]>0.
\end{displaymath}
Recall that the AD definition of conserved energy only makes sense 
in HJF (and EF) since we have no notion of total energy for 
fourth--order equations of motion. Total gravitational energy in 
Jordan frame is therefore defined as a quantity equal to that in HJF. 

\section{Minimum of the potential and stability} 
In order to establish whether the potential for the scalar gravity 
$\phi$ in Einstein frame is non--negative in a vicinity of a candidate 
ground state solution $\cal{M}$, dS or AdS, it is necessary to 
calculate the first and second derivative of $V$ at this state. To 
this end one first determines the derivative $\frac{\ud r}{\ud p}$ 
of the inverse function $R=r(p)$ to the definition of the scalar, 
$p=\frac{\ud f}{\ud R}$. It is equal to 
\begin{equation}
\frac{\ud R}{\ud p}=\left. \left(\frac{\ud p}{\ud r}\right)^{-1}
\right|_{r=r(p)}=\left[\left.\frac{\ud^2 f}{\ud R^2}\right|_{R=
r(p)}\right]^{-1}.
\end{equation}
Applying this outcome to the potential in (10) one finds 
\begin{equation}
\frac{\ud V}{\ud p}= \frac{1}{2\kappa^2 p^2}\left[\frac{2}{p}
f\left(r(p)\right)-r(p)\right]
\end{equation}
and this expression should also be inserted into the field equation 
(12) for $\phi$.\\

Consider a CGS solution in Jordan frame with $G_{\alpha\beta}=
-\frac{1}{4}\lambda\bar{g}_{\alpha\beta}$ and $R(\bar{g})=\lambda$ 
where $\lambda$ is a solution to (2). In HJF the scalar $p$ at this 
state is $p_0\equiv p(\lambda)=f'(\lambda)>0$. For the function 
$r(p)$ one has $r(p_0)=r(f'(\lambda))=\lambda$. Under the conformal 
map from HJF to EF the metric $\bar{g}_{\mu\nu}$ of the CGS solution is 
mapped to $\tilde{g}_{\mu\nu}=p_0 \bar{g}_{\mu\nu}=f'(\lambda)\bar{g}_
{\mu\nu}$ and the scalar $\phi$ is equal to $\phi_0=\sqrt{\frac{3}{2}}
\frac{1}{\kappa}\ln f'(\lambda)$. The Einstein tensor remains invariant 
under a constant rescaling of the metric, hence 
\begin{displaymath}
G_{\mu\nu}(\bar{g})=\tilde{G}_{\mu\nu}(\tilde{g})=-\frac{1}{4}\lambda 
\bar{g}_{\mu\nu}=-\frac{1}{4}\frac{\lambda}{f'(\lambda)}
\tilde{g}_{\mu\nu}
\end{displaymath}
and this allows one to define a cosmological constant in Einstein frame 
as 
\begin{displaymath}
\Lambda \equiv \frac{\lambda}{4f'(\lambda)}. 
\end{displaymath}
Thus $\cal{M}$, dS and AdS spaces in JF (and HJF) are respectively mapped 
onto $\cal{M}$, dS and AdS spaces in EF satisfying $\tilde{G}_{\mu\nu}
(\tilde{g})=-\Lambda \tilde{g}_{\mu\nu}$ and being the CGS solutions 
in Einstein frame. Physical excitations of the field $\phi$ in EF should 
be counted from its ground value $\phi_0$, i.e., are equal $\psi \equiv 
\phi - \phi_0$, then $p= f'(\lambda)\exp(\sqrt{\frac{2}{3}}\kappa\psi)$. 
The potential $V$ at $\phi=\phi_0$ is 
\begin{displaymath}
V(\phi_0)= \frac{1}{2\kappa^2 p_0^2}[\lambda f'(\lambda)-f(\lambda)] 
\end{displaymath}
and applying (2) it equals to
\begin{displaymath}
V(\phi_0)= \frac{1}{4\kappa^2}\frac{\lambda}{f'(\lambda)}=
\frac{\Lambda}{\kappa^2}. 
\end{displaymath}
The potential for the scalar excitation $\psi$ is then 
\begin{equation}
U(\psi)\equiv V(\phi)-V(\phi_0)= V(p(\phi)) - \frac{\Lambda}{\kappa^2}
\end{equation}
and vanishes for vanishing excitation, $U(0)=0$. The field equation (11) 
is now modified to (hereafter $\tilde{g}_{\mu\nu}$ denotes any 
dynamical metric in EF, not only the maximally symmetic CGS solutions) 
\begin{equation}
\tilde{G}_{\mu\nu}(\tilde{g})+\Lambda \tilde{g}_{\mu\nu}=  
\kappa^2\left[\psi_{,\mu}\psi_{,\nu} - \frac{1}{2}\tilde{g}_{\mu\nu}
\tilde{g}^{\alpha\beta}\psi_{,\alpha}\psi_{,\beta} -
\tilde{g}_{\mu\nu} U(\psi)\right].
\end{equation}
The first derivative of $U$ with respect to $\psi$ (or $p$) vanishes 
when $\frac{2}{p}f(r(p))-r(p)=0$ 
and this equation viewed as an equation for $r$ coincides with eq. (2). 
Hence $\frac{\ud U}{\ud \psi}=0$ only at the CGS solutions with 
$r(p_i)=\lambda_i= r(f'(\lambda_i))$, $i=1,\dots, n$. In other terms 
the equation $\lambda f'(\lambda)=2f(\lambda)$ determines all stationary 
points of $U$. At each of these points the potential $U_i(\psi)=V(\phi)-
\Lambda_i/\kappa^2$ vanishes provided $\Lambda_i =\lambda_i(4f'(\lambda_
i))^{-1}$. On the other hand $U(\psi)$ (for a fixed value of $\lambda$) 
may also vanish at some points $r_i$ different from the solutions 
$\lambda_i$ but these are not its stationary points; if such points exist 
the dominant energy condition is broken and a kind of (nonlinear) 
instability may develop.\\

The second derivative of the potential, determining its behaviour at a 
stationary point is, from (20) and (19), 
\begin{equation}
\frac{\ud^2 U}{\ud \psi^2}= \frac{1}{3p^2}\left[-4f(r)+pr+
\frac{p^2}{f''(r)}\right].
\end{equation}
At the stationary point $R=r=\lambda$, $\psi=0$ and $p=f'(\lambda)$, then 
\begin{equation}
\frac{\ud^2 U}{\ud \psi^2}\Big|_{\psi=0}= \frac{-\lambda}{3f'(\lambda)}+
\frac{1}{3f''(\lambda)}.
\end{equation}
For regular Lagrangians we are considering in this work one has 
$f''(\lambda)\ne 0$ finite. The potential $U(\psi)$ is non--negative if 
it attains minimum at $\psi=0$, i.e.  
\begin{displaymath}
\frac{\ud^2 U}{\ud \psi^2}\Big|_{\psi=0}>0.
\end{displaymath}
Stability conditions were also derived by other authors applying 
different methods. Our condition is equivalent to that found in 
\cite{Fa1} which after using $\lambda f'(\lambda)=2f(\lambda)$ reads  
\begin{equation}
3f'(\lambda) \frac{\ud^2 U}{\ud \psi^2}\Big|_{\psi=0}\geq 0. 
\end{equation}
The linear perturbation method applied in \cite{Fa1} implies that 
stability occurs whenever the weak inequality in (25) holds. 
Cognola et al. \cite{CGZ} employ a minisuperspace approach to the 
stability problem (perturbations are spatially homogeneous) and get 
a stability condition of de Sitter space which is 
equivalent\footnote{The formula for the condition seems to be 
misprinted since it disagrees with their earlier work.}
to ours; also computing one--loop quantization 
corrections to $L=f(R)$ they find this condition for dS background 
\cite{CZ}. Yet Song et al. \cite{SHS, SH} define stability of 
spatially flat R--W spacetime in a nonstandard way: a gravity theory 
is stable if it approaches general relativity at high curvatures (for 
small $R$ the theory should diverge from GR by definition); this 
cosmological criterion does not deal with a ground state solution.\\ 
The derivation of (24) holds both for $\lambda=0$ and $\lambda\ne 0$. 
The case $\lambda=0$ is simpler to study. In this case $f(0)=0$ and 
assuming analyticity around $R=0$ one has 
\begin{equation}
f(R)= R +aR^2 +\sum_{n=3}^{\infty} a_n R^n, 
\end{equation}
then $f(0)=0$, $f'(0)=1$, $f''(0)= 2a\ne 0$ and $U''(0)= \frac{1}{6a}$. 
For $a>0$ the potential $U\geq 0$ and the scalar field satisfies DEC. For 
spacetimes which are asymptotically flat it is known \cite{MS1} that 
$E_{ADM}[g]= E_{ADM}[\tilde{g}, \psi]\geq 0$ and the total energy vanishes 
only in Minkowski spacetime, $\tilde{g}_{\mu\nu}=\eta_{\mu\nu}=g_{\mu\nu}$ 
and $\psi=0$. \\

In de Sitter space ($\lambda>0$) there are following cases:\\
--- for $f''(\lambda)<0$ the potential attains maximum at $\psi=0$ and 
the space is unstable;\\
--- for $f''(\lambda)>0$ and $f'(\lambda)> \lambda f''(\lambda)$ there 
is minimum of $U$ and $U(\psi)\geq 0$, hence the space is stable;\\
--- for $f''(\lambda)>0$ and $f'(\lambda)< \lambda f''(\lambda)$ one 
finds $U''(0)<0$ and instabilities develop.\\
For anti--de Sitter space the situation is reversed:\\
--- for $f''(\lambda)<0$ and $f'(\lambda)> \lambda f''(\lambda)$ the 
negative potential attains maximum and the space is unstable;\\
--- for $f''(\lambda)<0$ and $f'(\lambda)< \lambda f''(\lambda)$ the 
potential is at minimum and AdS is stable;\\
--- for $f''(\lambda)>0$ the minimum of $U$ shows stability of the space.\\

Finally we return to the problem of singular Lagrangians (13) for which 
$f(\lambda)=0=f'(\lambda)$; for them the derivative (24) is divergent 
and the method of deriving it does not work. One may instead apply the 
gauge invariant perturbation method for de Sitter space directly in 
Jordan frame which gives rise \cite{Fa1} to the inequality (25). Let the 
lowest nonvanishing coefficient in the series (13) be $a_k$. If $k>2$ 
then also $f''(\lambda)=0$ and the expression (25) becomes indeterminate. 
In order to give it a definite value we define a function 
\begin{equation}
J(R) \equiv -R + \frac{f'(R)}{f''(R)}
\end{equation}
and define $J(\lambda)$ as its limit for $R\to \lambda$. Let $R= 
\lambda +\epsilon$, $|\epsilon|\ll 1$, then $f'(R)=k a_k\epsilon^{k-1}+
O(\epsilon^k)$, $f''(R)= k(k-1) a_k\epsilon^{k-2}+ O(\epsilon^{k-1})$ 
and 
\begin{displaymath}
J(\lambda +\epsilon)=-(\lambda +\epsilon)+
\frac{\epsilon}{k-1}+O(\epsilon^2). 
\end{displaymath}
Hence the stability criterion is $J(\lambda)=-\lambda\geq 0$. Recall 
that the method works only in dS space, $\lambda\geq 0$, therefore the 
conclusion is that for all NLG theories having Lagrangians of the form 
(13) with $\lambda>0$, de Sitter space (as a CGS 
solution\footnote{Besides $R=\lambda$ there are in general other 
solutions to $Rf'(R)=2f(R)$, e.g.~for $f(R)=a(R-\lambda)^3$ the other 
solution is $R=-2\lambda$.}) is unstable. None of the methods can be 
applied to these Lagrangians in the case $\lambda<0$. It might be 
argued that by continuity the criterion $J(\lambda)\geq 0$ should also 
work for $\lambda<0$, then all AdS spaces would be stable in these 
theories. However this argument is of little reliability.\\

In Paper I an astonishing theorem was mentioned to the effect that an 
anti--de Sitter space may be stable in spite of the fact that the 
scalar $\psi$ has a tachyonic mass (i.e., the potential $U(\psi)<0$ 
and attains maximum at this space) \cite{BF}. In fact, if small 
fluctuations of the scalar gravity vanish sufficiently fast at 
spatial infinity of AdS space (i.e., for $r\to \infty$ in the metric 
(17)), the kinetic energy of the field dominates over its negative 
potential energy and the total energy of the scalar, 
\begin{displaymath}
E(\psi)=-\int\,\ud^3 x\, \sqrt{-\bar{g}}\,T^{0\nu}\xi_{\nu},
\end{displaymath}
where $T^{0\nu}$ is given in (22), is finite and positive, $0<E(\psi)
<\infty$. This occurs if $\frac{\ud^2 U}{\ud \psi^2}>\frac{3}{4}
\Lambda$ at $\psi=0$. Since the energy of gravitational perturbations 
of AdS space is positive \cite{AbD}, the total energy of metric and 
scalar field fluctuations is positive and Breitenlohner and Freedman 
conclude \cite{BF} that AdS space is stable against these (small) 
fluctuations. Applying the definition of $\Lambda$ in Einstein frame 
arising in NLG theories, the condition of stability of AdS in the case 
of maximum of the potential reads 
\begin{equation}
0> \frac{\ud^2 U}{\ud \psi^2}\Big|_{\psi=0}> \frac{3\lambda}
{16f'(\lambda)}.
\end{equation}
It should be stressed, however, that in this case the DEC  
is violated (only the total energy of the scalar is positive). 
From the viewpoint of a rigorous mathematical approach to the stability 
problem the condition (28) is rather unreliable \cite{An3}. 

\section{Examples: specific Lagrangians}
We now apply the stability criteria of the previous section to a 
number of Lagrangians, some of which were already discussed in the 
literature. We assume that the Lagrangians depend on one dimensional 
constant $\mu$ and some dimensionless constants. $\mu$ is positive and 
has dimension 
$(\textrm{length})^{-1}$ so that $R/\mu^2$ is a pure number.\\

\begin{equation}
1. \qquad L=R+\frac{\mu^{4n+4}}{R^{2n+1}}, \quad n=0,1,\ldots.
\end{equation}
This Lagrangian belongs to the class which admits no CGS solutions 
since it is given by eq. (3) for $a=0$ and $F(R) =-R-(2n+3)\mu^
{4n+4}R^{-(2n+1)}$. Clearly it should be rejected.
Yet according to Sawicki and Hu \cite{SH} the theory for $n=0$ 
converges to general relativity for large $R$ and in this sense 
is admissible.\\

\begin{equation}
2. \qquad L=R+\frac{\mu^{4n+2}}{R^{2n}}, \quad n=1, 2,\ldots.
\end{equation}
There is only one CGS solution with $\lambda=\lambda_{-}\equiv 
-(2n+2)^{\frac{1}{2n+1}}\mu^2 <0$, $f'(\lambda)=(2n+1)(n+1)^{-1}>0$ 
and the scalar is
\begin{equation}
p(r)=1-2n\left(\frac{\mu^2}{r}\right)^{2n+1}.
\end{equation}
We consider spacetimes with $R=r$ in vicinity of $R=\lambda_{-}$, 
so that $-\infty<r<0$ and $1<p<+\infty$. The inverse function and 
the potential are respectively 
\begin{equation}
r = -\left(\frac{2n}{p-1}\right)^{\frac{1}{2n+1}} \mu^2, \quad
U=U_{-}=-\frac{(2n+1)\mu^2}{2\kappa^2 p^2}
\left(\frac{p-1}{2n}\right)^{\frac{2n}{2n+1}} -
\frac{\Lambda_{-}}{\kappa^2}
\end{equation}
with $\Lambda_{-}<0$. The potential is always non-negative and 
$U\leq \left|\frac{\Lambda_{-}}{\kappa^2}\right|$. It attains minimum 
at $\psi=0$ showing that AdS space is a stable ground state solution 
for this theory.\\

\begin{equation}
3. \qquad L=R+\frac{\mu^{2n+2}}{R^n} \quad \textrm{for} \quad
-1<n<0 \quad \textrm{real},
\end{equation}
for non--integer $n$ the function $R^n$ is replaced by $|R|^n$.  
The background evolution of the R--W spacetime is cosmologically acceptable 
\cite{AGPT} and solutions of the 
linear perturbation equations for this Lagrangian are not incompatible 
with the observational data \cite{LiB}. However the equation for a 
ground solution gives rise to the contradiction 
$|R|^{n+1}=-(n+2)\mu^{2n+2}$ implying that the theory in untenable.\\

\begin{equation}
4. \qquad L=R-\frac{\mu^{4n+2}}{R^{2n}}, \quad n=1, 2,\ldots.
\end{equation}
Here $\lambda$ and $r(p)$ have the same moduli and opposite sign to 
those in the case 2: dS space is the unique CGS solution for 
$\lambda=\lambda_{+}=-\lambda_{-}>0$ and $r(p)>0$. 
Accordingly, $\Lambda=\Lambda_{+}=-\Lambda_{-}$. Now we take $r$ 
around $r=\lambda_{+}$, and again $1<p<\infty$. The potential is 
$U=U_{+}=-U_{-}$, hence it is contained in the interval 
$-\frac{\Lambda_{+}}{\kappa^2}\leq U_{+}\leq 0$. This indicates that 
$U$ has maximum at 
$\psi=0$ and this fact is confirmed by a direct computation. In conclusion, 
de Sitter space is unstable and this theory is discarded as unphysical.\\

\begin{equation}
5. \qquad L=R-\frac{\mu^{4n+4}}{R^{2n+1}}, \quad n=0, 1,\ldots.
\end{equation}
This Lagrangian has been most frequently studied in applications to the 
accelerating universe, usually for $n=0$. Most expressions here are 
akin to respective ones in the case 4. The field $p$ is always greater 
than 1 and there are two CGS solutions for 
$\lambda_{\pm}=\pm(2n+3)^{\frac{1}{2n+2}} \mu^2$,
hence $p(\lambda_{\pm})=p(\lambda_{-})=\frac{4n+4}{2n+3}$. The two CGS 
solutions define two different sectors of the theory which should be 
separately studied.\\
A. De Sitter space sector.\\
$\lambda=\lambda_{+}>0$ and the sector comprises all positive values of $r$. 
The inverse function is 
\begin{equation}
r(p)=r_{+}(p)=\left(\frac{p-1}{2n+1}\right)^{\frac{-1}{2n+2}}
\mu^2
\end{equation}
giving rise to the potential \cite{Carr1}
\begin{equation}
U(p(\psi))=U_{+}=\frac{n+1}{\kappa^2 p^2}
\left(\frac{p-1}{2n+1}\right)^{\frac{2n+1}{2n+2}}\mu^2 -
\frac{\Lambda_{+}}{\kappa^2}
\end{equation}
which is always non-positive and attains maximum at dS space. This space 
is then unstable (for $n=0$ it was found in \cite{Carr1, Fa1, CZ}) and 
this sector of the theory must be rejected (on other grounds this 
conclusion was derived in \cite{SW}).\\
Seifert \cite{Sei} finds that gravity theory (35) is highly unstable in 
the presence of matter: a static spherically symmetric solution 
becomes unstable to linear spherically symmetric perturbations if 
perfect fluid matter forms a quasi-Newtonian polytropic star. This result 
is derived applying an intricate variational method and requires very 
long computations. We note that (besides the fact that the Newtonian 
limit is not well defined there) the author assumes that $R$ is 
approximately equal to the stellar matter density. This means that he 
deals with spherically symmetric perturbations of dS space. Since this 
space is unstable in pure gravity (35) it would be rather surprising if 
a small amount of matter could stabilize it.\\

B. Anti--de Sitter space sector.\\
Its existence (for $n=0$) was first noticed in \cite{Carr1}, then in 
\cite{CENOZ}, but its properties were 
never analyzed in detail, probably due to the fact that a negative 
$\Lambda$ does not fit the observed accelerated expansion. $\lambda=
\lambda_{-}<0$ and accordingly $-\infty<r<0$, hence $r(p)=r_{-}(p)=
-r_{+}(p)$ and $U=U_{-}=-U_{+}$ with $\Lambda_{-}=-\Lambda_{+}$. This 
potential is non-negative and has minimum at $\psi=0$. This sector 
has a stable ground state solution\footnote{In \cite{Fa2} it is claimed 
that Lagrangians given in cases 4 and 5 (for both $n$ even and odd) 
always develop instabilities while Lagrangians in cases 1 and 2 always 
describe a stable theory.} and in this sense it forms a viable gravity 
theory. The scalar gravity has mass being a function of $n$, for $n=0$ 
it is $m^2=\frac{3\sqrt{3}}{4}\mu^2$ while for $n\to\infty$ it tends to 
$m^2\to\frac{\mu^2}{6}$. Disregarding the incompatibility of this 
theory with the cosmic acceleration, one may make a rough estimate of 
$\mu$. Since $\Lambda$ is of order $-\mu^2$ for all $n\geq 0$ and the 
observational limit is $|\Lambda|\leq 10^{-52}\textrm{m}^{-2}$ one gets 
an upper limit $\mu\leq 10^{-26}\textrm{m}^{-1}$ or $\mu\hbar c\leq 
10^{-33}\textrm{eV}$, very small indeed. \\

\begin{equation}
6. \qquad L=R \left(\ln\frac{R}{\mu^2}\right)^q, 
 \quad q>0.
\end{equation}
One assumes $R>0$. A unique possible ground state is dS space with 
$\lambda = e^q \mu^2$, then $p(\lambda)= 2\exp (q\ln q)$ and 
$f''(\lambda) = \mu^{-2}(2q -1)\exp[(q-1)\ln q -q]$. The Lagrangian 
must be regular, i.e.~$f''(\lambda)\neq 0$ implying $q\neq 1/2$. 
From (24)
\begin{displaymath}
U''(0)=\frac{\mu^2}{6(2q-1)}e^{q(1-\ln q)}
\end{displaymath}
hence for $0<q< 1/2$ dS space is unstable while for $q>1/2$ de Sitter 
space is a stable ground state. The function $p(r)$ may be inverted 
and then the potential can be explicitly calculated only for $q=1$ 
or 2. According to \cite{AGPT} this theory is cosmologically acceptable 
for any $q>0$ though the matter era begins too early and its duration 
is too long.\\

\begin{equation}
7. \qquad L=\alpha R-\frac{\mu^2}{\sinh\frac{R}{\mu^2}}, \quad 
\alpha\geq 0.
\end{equation}
This Lagrangian appeared in the metric--affine approach to gravity 
\cite{ABF}. The equation $Rf'(R)-2f(R)=0$ cannot be analytically solved 
even in the case $\alpha=0$ (it can only be shown that the roots do not 
lie close to $R=0$) and for practical reasons this theory must be 
rejected.\\

\begin{equation}
8. \qquad L=\mu^2 \left(\ln\frac{R}{\mu^2}+\frac{1}{2}\right)
+\frac{a}{\mu^2}R^2, \quad a>0.
\end{equation}
One may start from a more general Lagrangian \cite{NO3}
\begin{displaymath}
L=\gamma R+b\ln(cR)+\frac{a'}{\mu^2}R^2, \qquad a', b, c>0,
\end{displaymath}
but then eq. (2) for $\lambda$ cannot be solved analytically. We 
therefore set $\gamma=0$ and multiply $L$ by $\mu^2/b$ and define 
$a=\frac{a'}{b}\mu^2$; finally we choose such value of $c$ as to get a 
simple expression for $\lambda$. A unique solution to (2) is then 
$\lambda=\mu^2$ and 
\begin{equation}
p=\frac{\mu^2}{r}+\frac{2a}{\mu^2}r,
\end{equation}
$r>0$. To invert this function we first notice that $p(r)\to\infty$ 
for both $r\to 0$ and $r\to\infty$ and has minimum at $r_0=\mu^2/
\sqrt{2a}$ equal to $p(r_0)=2\sqrt{2a}$. Hence $p(r)$ may be inverted 
either in the interval $0<r<r_0$ or $r>r_0$. To choose the correct 
interval one must establish whether $\lambda=\mu^2$ belongs to the 
ascending or descending branch of $p(r)$ and this depends on the value 
of $a$. We assume $a>1/2$, then $\mu^2>r_0$ and dS space lies on the 
ascending branch of $p$ (for $a<1/2$ a similar procedure can be 
performed). Solving (41) one chooses the larger root (both roots are 
positive), 
\begin{equation}
r(p)=\frac{\mu^2}{4a}\left(p+\sqrt{p^2-8a}\right)
\end{equation}
since $r\to\infty$ corresponds to $p\to\infty$. The potential is 
\begin{equation}
U=\frac{1}{16a\kappa^2 p}
\left[P(p)-\frac{8a}{p}\left(\ln P-\ln(4a)\right)\right]\mu^2 -
\frac{\Lambda}{\kappa^2},
\end{equation}
where $P\equiv p+\sqrt{p^2-8a}$ and $\Lambda=\frac{\mu^2}{4(2a+1)}$. 
This implies $f'(\lambda)=2a+1>\lambda f''(\lambda)=2a-1$ and the 
potential has minimum at $\psi=0$. This theory has dS space as a stable 
ground state and is viable.\\
The case $a=1/2$ is singular since $f''(\mu^2)=0$ and $p(r)$ cannot be 
inverted around $r=\mu^2$ while $f'(\mu^2)=2$. Formally the conformal 
map to EF exists at this point but the potential $U$ cannot be defined 
there. None of the methods to check the stability does work there 
and it is reasonable to disregard this case.\\

9. The limiting case $a=0$ of the Lagrangian (40)  
requires a separate treatment. Again $\lambda=\mu^2$, $p=\frac{\mu^2}{r}$, 
and $r(p)=\mu^2/p>0$. $f'(\lambda)=1$ 
and $f''(\lambda)=-\frac{1}{\mu^2}$ give rise to $U''(0)=-\frac{2}{3}
\mu^2$. De Sitter space is unstable making the theory untenable.\\
The additive constant appearing in this Lagrangian (as well as in the 
case 8) is inessential in the sense that it only affects the absolute 
value of $\lambda$ (but not its sign) and has no influence on stability 
properties of dS space. In fact, for a Lagrangian 
\begin{displaymath}
 L=\mu^2 \left(\ln\frac{R}{\mu^2}+a\right),
\end{displaymath}
$a$ real dimensionless, one gets again $p=\frac{\mu^2}{r}$ and the value 
of $\lambda$ is shifted to $\lambda=\mu^2 \exp(\frac{1}{2}-a)$, hence it is 
still dS space. Then  
$U''(0)=-\frac{\mu^2}{3}\left[1+\exp(1-2a)\right]$ implying instability 
of the space for any $a$. This case is, however, exceptional: 
we will see below that in general not only $\mu$ but also dimensionless 
parameters in $L$ determine stability of CGS solutions.\\

\begin{equation}
10. \qquad L=\mu^2 \left(\frac{R}{\mu^2}\right)^{\alpha}
\end{equation}
for $\alpha$ rational (negative and positive) has also attracted some 
attention \cite{Fa1, ABF, CB1} since it is a scale--invariant theory. 
For non--integer $\alpha$ one takes $|R|^{\alpha}$. If 
$\alpha<0$ the equation $Rf'(R)=2f(R)$ is solved only by 
$R=\pm\infty$ and we reject this case. For $\alpha=2$ one gets the 
degenerate Lagrangian $R^2$ which we disregard. For $\alpha>2$ integer 
this is a singular Lagrangian (13) discussed in section 4 having 
$\lambda=0$ and the criterion $-\lambda\geq 0$ yields that Minkowski 
space is stable for these theories. Putting aside the obvious case 
$\alpha=1$ one considers $\alpha>0$ non--integer. $f(0)=0$ always. 
For $0<\alpha<1$ both $f'(0)$ and $f''(0)$ are infinite, for 
$1<\alpha<2$ there is $f'(0)=0$ and $f''(0)=\infty$ and for 
$\alpha>2$ both $f'(0)=f''(0)=0$. Once again one may apply the function 
$J(R)$ defined in (27) and it is equal $J=\frac{2-\alpha}{\alpha-1}R$ 
so that $J(0)=0$ and for all three cases the criterion $J(0)\geq 0$ 
is satisfied. One may thus claim that for all $\alpha>0$ Minkowski 
space is the unique stable ground state, nevertheless it is difficult 
to avoid impression that for $\alpha\ne 1$ the theory is bizarre and 
rather unphysical (and furthermore in conflict with the astronomical 
observations, as mentioned in Paper I).\\

\begin{equation}
11. \qquad L=R-\frac{\mu^4}{R}+\frac{a}{\mu^2}R^2, 
\end{equation}
$a$ real \cite{NO1, Di, Fa1}. There 
are two CGS solutions with $\lambda_{\pm}=\pm\sqrt{3}\mu^2$, 
which are the same as for the case $a=0$ (Lagrangian (35) for 
$n=0$) since the $R^2$ term does not contribute to $\lambda$. The 
attempt to find the inverse function $r(p)$ leads to a cubic equation 
and solving it would be impractical. We therefore quit from computing 
the explicit form of the potential (an implicit form of $V$ is given 
in \cite{NO1}) and restrict ourselves to studying its extrema.\\
A. De Sitter sector for $\lambda=\lambda_{+}$.\\
The condition $p(\lambda_{+})>0$ requires $a>-\frac{2}{3\sqrt{3}}$. 
This condition does not determine the sign of 
\begin{displaymath}
f''(\lambda_{+}) =\frac{2}{\mu^2}\left(\frac{-1}{3\sqrt{3}}+a\right)
\end{displaymath}
and from (24) one finds:\\
--- for $-\frac{2}{3\sqrt{3}}<a<\frac{1}{3\sqrt{3}}$ dS space is 
unstable and \\
--- for $a>\frac{1}{3\sqrt{3}}$ dS space is stable. Yet cosmologically 
the theory in this case is inacceptable since there is no standard matter 
era preceding the acceleration era \cite{AGPT}.\\
We omit the singular case $a=\frac{1}{3\sqrt{3}}$ where $f''(\lambda_{+})
=0$.\\
 B. Anti--de Sitter sector with $\lambda=\lambda_{-}$.\\
 Now the condition $p(\lambda_{-})>0$ requires $a<\frac{2}{3\sqrt{3}}$. 
 From (24) one gets that for  $-\frac{1}{3\sqrt{3}}<a<\frac{2}{3\sqrt{3}}$ 
 the potential has minimum at $\psi=0$ and AdS space is stable. Yet for 
 $a<-\frac{1}{3\sqrt{3}}$ the potential has maximum. This, however, does 
 not automatically imply the instability since one should furthermore 
 apply the criterion (28) of positivity of scalar field energy. It follows 
 from it that AdS space is\\
 --- stable for $-\frac{19}{9\sqrt{3}}<a<-\frac{1}{3\sqrt{3}}$ with respect 
 to scalar field perturbations with positive energy,\\
 --- unstable for $a<-\frac{19}{9\sqrt{3}}$.\\
 In the range of values of $a$ for which the theory is stable in the 
 standard sense (the potential has minimum) the mass of the scalar 
 gravity excitations above dS space is
 \begin{displaymath}
m_{+}^2 =3\sqrt{3}[2(3\sqrt{3}a+2)(3\sqrt{3}a-1)]^{-1}\mu^2,
\end{displaymath}
while in the case of AdS ground state it is
\begin{displaymath}
m_{-}^2 =3\sqrt{3}[2(3\sqrt{3}a+1)(2-3\sqrt{3}a)]^{-1}\mu^2.
\end{displaymath}
The particle masses tend to infinity when $a$ approaches the finite limits 
of the admissible range. $m_{+}$ monotonically decreases and becomes very 
small for large values of $a$ while in the AdS sector the scalar particle mass 
attains minimum $m_{-}^2=\frac{2}{\sqrt{3}}\mu^2$ at $a=(6\sqrt{3})^{-1}$. \\
In the interval $\frac{1}{3\sqrt{3}}<a<\frac{2}{3\sqrt{3}}$ the theory has 
two viable sectors: one with dS space ground state for 
$\Lambda_{+}=3\sqrt{3}[8(3\sqrt{3}a+2)]^{-1}\mu^2$ and the other having 
AdS as a ground state with $\Lambda_{-}=-3\sqrt{3}[8(2-3\sqrt{3}a)]^{-1}
\mu^2$. Classically these are two different physical theories, each with 
a unique ground state. One cannot claim that this is one theory having two 
different and distant (in the space of solutions) local minima of energy. 
Energetically these two states are incomparable, each of them has vanishing 
energy (defined with respect to itself) and assuming that one of these 
minima is lower than the other is meaningless \cite{Wi}. One may only 
compare the masses of the scalar gravity in the two theories. The mass 
ratio $\left(\frac{m_{+}}{m_{-}}\right)^2$ decreases monotonically from 
infinity for $a$ approaching $(3\sqrt{3})^{-1}$ to zero for $a$ tending 
to $2(3\sqrt{3})^{-1}$. If one believes that this Lagrangian describes the 
physical reality a difficult problem arises: how does the nature choose 
which of the two theories with the same Lagrangian is to be realized? In 
our opinion the nature avoids this problem merely by avoiding this 
Lagrangian (and other ones with the same feature).\\
This Lagrangian illustrates a general rule: all the parameters appearing 
in a Lagrangian do contribute to determination of stable sectors (i.e., 
physically distinct theories) corresponding to it.\\

\begin{equation}
12. \qquad L= R\exp\left(\frac{\theta \mu^2}{R}\right), 
\qquad \theta=\pm 1.
\end{equation} 
 
To each value of $\theta$ there is one CGS 
solution with $\lambda=-\theta \mu^2$ and $p(\lambda)=2/e>0$, then
$U''(0)=-\frac{1}{6} \theta e\mu^2$. For $\theta=-1$ the potential 
attains minimum and de Sitter space is classically stable. For 
$\theta=+1$ one applies the stability criterion (28) for AdS space 
and one gets that this solution is unstable according to this 
condition too. The Lagrangian (46) for $\theta =-1$ is also 
cosmologically preferred since it is asymptotically equivalent to 
the $\Lambda$CDM model \cite{AGPT}. Unfortunately the function $p(r)$ 
cannot be inverted analytically and the explicit form of the potential 
is unavailable.\\

13. Finally we consider a class of "toy models"' possessing infinite 
number of ground states. For convenience we introduce a dimensionless 
variable $x=R/\mu^2$ and assume
\begin{equation}
L=f(R)=\mu^2 F(x)=\mu^2 e^{2I(x)}
\end{equation}
where 
\begin{equation}
I(x)\equiv \int\frac{\ud x}{x+h(x)}
\end{equation}
and $h(x)$ is a continuous periodic function taking both positive and negative 
values, $M_1\leq h(x) \leq M_2$ with $M_1<0$ and $M_2>0$. The scalar field is 
\begin{displaymath}
p=\frac{\ud F}{\ud x}=\frac{2}{x+h(x)}F(x)
\end{displaymath} 
and is positive if $x+h(x)>0$. For an arbitrary $h$ one cannot find $r(p)$ 
and the potential; here it is sufficient to determine CGS solutions and 
$U''$ at these states. The eq. (2) takes now the form $x\frac{\ud F}{\ud x}=
2F$ and since $F>0$ it is equivalent to
\begin{equation}
x=\frac{2}{\frac{\ud}{\ud x}\ln F}.
\end{equation}
On the other hand from the definitions (47) and (48) it follows that 
\begin{equation}
\frac{2}{\frac{\ud}{\ud x}\ln F}=x+h(x),
\end{equation}
hence those $x$ which are solutions of (49) must also be solutions to $h(x)
=0$. Since $M_1\leq h(x) \leq M_2$ there is at least one root of $h(x)=0$ 
and for a continuous periodic function there is infinite number of zeros, 
$h(x_n)=0$, $n=0, 1,\ldots$ and $\lambda_n=\mu^2 x_n$. Note that $x_n\neq 
0$ since $\lambda=0$ implies $f(0)=\mu^2\exp(2I(0))=0$ while $I(0)$ is finite 
by its definition. The function $x+h(x)$ tends to $\pm\infty$ for $x\to
\pm\infty$, hence there is a point $x=y$ such that $y+h(y)=0$ and $y\neq 0$. 
To ensure that $x+h(x)>0$ for $x>y$ one requires $x+h(x)$ be monotonic, i.e.~ 
$1+h'(x)>0$. Then $I(x)$ is defined (and positive) for all $x>y$. Denoting 
$I_n\equiv I(x_n)$ one finds that $U''$ at a point $R=\lambda_n$ is 
\begin{equation}
\frac{\ud^2 U}{\ud \psi^2}\Big|_{\lambda_n}= \frac{\mu^2}{6}x_n^2
e^{-2I_n}\frac{h'(x_n)}{1-h'(x_n)}.
\end{equation}
The condition $h'(x)>-1$ does not determine the sign of the fraction and to 
this aim one must specify $h$. Here we choose as an example $h(x)\equiv
\frac{1}{2}(\sin x-\cos x)$. Clearly $h'(x)=\frac{1}{2}(\sin x+\cos x)>-1$ 
and the unique solution of $x+\frac{1}{2}(\sin x-\cos x)=0$ is 
$y=0,3183\ldots$. The zeros $x_n>y$ of $h$ are solutions to 
$\textrm{tg}\,x=1$ ($\cos x\neq 0$) and these are $x_n=\frac{\pi}{4}+n\pi$, 
$n=0, 1, \ldots$. At these points $h'(x_n)=(-1)^n\frac{\sqrt{2}}{2}$ and for 
$n$ odd there is 
\begin{displaymath}
\frac{\ud^2 U}{\ud \psi^2}\Big|_{\lambda_{2n+1}}<0,
\end{displaymath} 
therefore the infinite sequence of dS spaces with curvatures $\lambda_{2n
+1}=\mu^2 x_{2n+1}$ defines unphysical (unstable) sectors of the theory. 
Yet the other sequence for $n$ even consists of dS spaces having curvatures 
$\lambda_{2n}=(2n+\frac{1}{4})\pi\mu^2$ which are stable for this 
Lagrangian. The scalar particles corresponding to these sectors have 
masses 
\begin{displaymath}
m_{2n}^2=\frac{(2n+\frac{1}{4})^2}{6(\sqrt{2}-1)}\pi^2\mu^2
e^{-2I_{2n}}.
\end{displaymath}

\section{Conclusions}
In this paper we have investigated stability of ground state solutions in 
$L=f(R)$ gravity theories being either Minkowski, de Sitter or anti--de 
Sitter spaces. Stability may be studied in any frame and Einstein frame is 
particularly suitable to this aim since one may apply there the methods 
developed in general relativity. We have given an explicit, effective and 
simple method of checking stability of these spaces based on the dominant 
energy condition applied to the 
scalar component of the gravitational doublet. After applying the method 
to thirteen specific Lagrangians (their ground states are de Sitter 
and/or anti--de Sitter spaces) corresponding to 20 different cases 
(depending on values of parameters in $L$) it was found that, as it was 
a priori expected, half of them give rise to viable theories (9 viable
versus 11 untenable ones). And 
a generic feature is the existence of multiple vacua (stable ground states), 
each generating a separate physical sector or rather a separate gravity 
theory, all having the same Lagrangian. Hence it is expected that there is 
an infinity of viable gravity theories. What to do with such a wealth of 
theories (all differing from each other only by the form of the potential 
for the scalar gravity field)? \\

We stress that it is incorrect merely to search for a theory which easily 
and immediately accounts for the big problem of cosmology---the 
acceleration of the universe. After all general relativity was not 
formulated to solve some urgent problems in celestial mechanics (the 
perihelion shift of Mercury) or in cosmology (non--existence of Newtonian 
cosmology) and for many years its confirmation was quite marginal. At 
the time of its advent its advantage was that it was physically much 
deeper and more general than Newton's gravity. And the same should be 
expected about a modified gravity which may ultimately replace Einstein's 
theory. Its physical content will be more relevant than immediate 
observational confirmation. \\
Before a deep creative physical idea will appear we need further viability 
criteria to maximally reduce the set of viable gravity theories. 
Undoubtedly one of the most important 
ones will be the condition that a tenable theory must be in agreement with 
the Newtonian and post--Newtonian approximations to gravity---as soon as 
these approximations will be rigorously defined in de Sitter and anti--de 
Sitter backgrounds. One should, however, expect that the selected set 
will still be large and possibly infinite. 

\section*{Acknowledgments}
 
I am grateful to Michael Anderson, Piotr Bizo\'n, Piotr Chru\'sciel, 
Helmut Friedrich, Zdzis\l{}aw Golda and Barton Zwiebach for extensive 
discussions, helpful comments and explanations.
This work is supported in part by a Jagellonian University grant.


\begin{thebibliography}{}
\frenchspacing
\bibitem{Sok}
L. M. Soko\l{}owski, \textit{Metric gravity theories and cosmology. 
I. Physical interpretation and viability}, Class. Quantum Grav. 
\textbf{24} (2007) 3391 [gr-qc/0702097]. 


\bibitem{AGPT}
L. Amendola, R. Gannouji, D. Polarski and S. Tsujikawa, 
Phys. Rev. \textbf{D75}: 083504, 2007  [gr-qc/0612180].

\bibitem{BRu}
K. A. Bronnikov and S. G. Rubin, 
Phys. Rev.  \textbf{D73} (2006) 124019; 
K. A. Bronnikov, R. V. Konoplich and S. G. Rubin,
Class. Quantum Grav. 
\textbf{24} (2007) 1261 [gr-qc/0610003]. 

\bibitem{dHJ}
E. d'Hoker and R. Jackiw, 
Phys. Rev.  \textbf{D26} (1982) 3517. 

\bibitem{Wi}
E. Witten, 
Nucl. Phys. \textbf{B195} (1982) 481. 

\bibitem{Woo}
R. P. Woodard, \textit{Proceedings of 3rd Aegean Summer School 
"The Invisible Universe: Dark Matter and Dark Energy", 
September 2005}, [astro-ph/0601672]. 

\bibitem{AbD}
L. F. Abbott and S. Deser, 
Nucl. Phys. \textbf{B195} (1982) 76. 

\bibitem{GHW}
G. W. Gibbons, C. M. Hull and N. P. Warner, 
Nucl. Phys. \textbf{B218} (1983) 173. 

\bibitem{HM}
G. T. Horowitz and R. C. Myers, 
Phys. Rev.  \textbf{D59} (1998) 026005. 

\bibitem{GP}
P. Ginsparg and M. J. Perry, 
Nucl. Phys. \textbf{B222} (1983) 245. 

\bibitem{ChK}
D. Christodoulou and S. Klainerman, \textit{The global nonlinear 
stablity of Minkowski space}, Princeton Univ. Press, Princeton 1993; 
S. Klainerman and F. Nicolo, \textit{The evolution problem in general 
relativity}, Birkh\"auser Verlag, Boston 2003. 

\bibitem{Zi}
N. Zipser, \textit{The global nonlinear stability of the trivial 
solution of Einstein--Maxwell equations}, Ph. D. thesis, Harvard 
University 2000, unpublished. 

\bibitem{LR}
H. Lindblad and I. Rodnianski, [math.AP/0411109]. 

\bibitem{Fr1}
H. Friedrich, 
J. Geom. Phys. \textbf{3} (1986) 101; Commun. Math. Phys. \textbf{107} 
(1986) 587.  

\bibitem{An1}
M. T. Anderson, 
Ann. H. Poincare \textbf{6} (2005) 801.

\bibitem{Fr2}
H. Friedrich, 
J. Diff. Geom. \textbf{34} (1991) 275. 

\bibitem{HR}
J. M. Heinzle and A. D. Rendall, Commun. Math. Phys. \textbf{269} 
(2007) 1, [gr-qc/0506134]. 

\bibitem{IW}
A. Ishibashi and R. M. Wald, 
Class. Quantum Grav. \textbf{21} (2004) 2981. 

\bibitem{Fr4}
H. Friedrich, 
J. Geom. Phys. \textbf{17} (1995) 125. 

\bibitem{An3}
M. T. Anderson (2006), private communication.

\bibitem{Ho}
G. T. Horowitz, 
Phys. Rev.  \textbf{D21} (1980) 1445. 

\bibitem{TH}
J. Traschen and C. T. Hill, 
Phys. Rev.  \textbf{D33} (1986) 3519. 

\bibitem{Fa1}
V. Faraoni, 
Phys. Rev.  \textbf{D72} (2005) 061501;
V. Faraoni and S. Nadeau, 
Phys. Rev.  \textbf{D72} (2005) 124005. 

\bibitem{Di}
R. Dick, 
Gen. Rel. Grav. \textbf{36} (2004) 217. 

\bibitem{NO1}
S. Nojiri and S. D. Odintsov, Phys. Rev. \textbf{D68} (2003) 123512. 

\bibitem{BO}
J. D. Barrow and A. C. Ottewill, 
J. Phys. \textbf{A16} (1983) 2757.

\bibitem{HOW1}
A. Hindawi, B. A. Ovrut and D. Waldram, 
Phys. Rev.  \textbf{D53} (1996) 5597. 

\bibitem{DK}
A. D. Dolgov and M. Kawasaki, 
Phys. Lett. \textbf{B573} (2003) 1. 

\bibitem{Sot}
T. P. Sotiriou, 
Phys. Lett. \textbf{B645} (2007) 389. 

\bibitem{Bi}
G. V. Bicknell, 
J. Phys. \textbf{A7} (1974) 1061. 

\bibitem{CB2}
T. Clifton and J. D. Barrow, 
Phys. Rev.  \textbf{D72} (2005) 123003. 

\bibitem{MFF1}
G. Magnano, M. Ferraris and M. Francaviglia, 
Gen. Rel. Grav. \textbf{19} (1987) 465. 

\bibitem{JK}
A. Jakubiec and J. Kijowski, 
Phys. Rev.  \textbf{D37} (1988) 1406; Gen. Rel. Grav. \textbf{19} 
(1987) 719. 

\bibitem{MS1}
G. Magnano and L. M. Soko\l{}owski,
Phys. Rev. \textbf{D50} (1994) 5039, [gr-qc/9312008].

\bibitem{HeW}
I. P. C. Heard and D. Wands, Class. Quantum Grav. 
\textbf{19} (2002) 5435, [gr-qc/0206085].

\bibitem{Sei}
M. D. Seifert, [gr-qc/0703060].

\bibitem{BGH}
W. Boucher, G. W. Gibbons and G. T. Horowitz, 
Phys. Rev. \textbf{D30} (1984) 2447.

\bibitem{CGZ}
G. Cognola, M. Gastaldi and S. Zerbini, 
[gr-qc/0701138].

\bibitem{CZ}
G. Cognola and S. Zerbini, 
J. Phys. \textbf{A39} (2006) 6245.

\bibitem{SHS}
Y.-S. Song, W. Hu and I. Sawicki, 
Phys. Rev. \textbf{D75} (2007) 044004 [astro-ph/0610532].

\bibitem{SH}
I. Sawicki and W. Hu, 
Phys. Rev. \textbf{D75} (2007) 127502 [astro-ph/0702278].

\bibitem{BF}
P. Breitenlohner and D. Z. Freedman, 
Phys. Lett. \textbf{115B} (1982) 197; Ann. Phys. (N.Y.) 
\textbf{144} (1982) 249. 

\bibitem{LiB}
B. Li and J. D. Barrow, 
Phys. Rev. \textbf{D75}: 084010, 2007  [gr-qc/0701111].

\bibitem{Carr1}
S. M. Carroll, V. Duvvuri, M. Trodden and M. S. Turner, Phys. Rev. 
\textbf{D70} (2004) 043528. 

\bibitem{SW}
M. E. Soussa and R. P. Woodard, 
Gen. Rel. Grav. \textbf{36} (2004) 855.

\bibitem{CENOZ}
G. Cognola, E. Elizalde, S. Nojiri, S. D. Odintsov and S. Zerbini,  
JCAP \textbf{0502} (2005) 010 [hep-th/0501096]. 

\bibitem{Fa2}
V. Faraoni, 
Phys. Rev.  \textbf{D74} (2006) 104017. 

\bibitem{ABF}
G. Allemandi, A. Borowiec and M. Francaviglia, Phys. Rev. 
\textbf{D70} (2004) 043524.

\bibitem{NO3}
S. Nojiri and S. D. Odintsov,
Gen. Rel. Grav. \textbf{36} (2004) 1765. 

\bibitem{CB1}
T. Clifton and J. D. Barrow, 
Phys. Rev.  \textbf{D72} (2005) 103005. 

\end{thebibliography}
\end{document}